# Physically-Based Particle Simulation and Visualization of Pastes and Gels

Claire Guilbaud, Annie Luciani, Nicolas Castagne
ACROE/ICA, INPG, Grenoble, France
{guilbaud, luciani, castagne}@imag.fr

## Abstract

This paper is focused on the question of simulation and visualization of 3D gel and paste dynamic effects.

In a first part, we introduce a 3D physically based particle (or mass-interaction) model, with a small number of masses and few powerful interaction parameters, which is able to generate the dynamic features of both gels and pastes. This model proves that the 3D mass-interaction method is relevant for the simulation of such phenomena, without an explicit knowledge of their underlying physics.

In a second part, we expose an original rendering process, the *Flow Structuring Method* that enhances the dynamic properties of the simulation and offers a realistic visualization. This process ignores all the properties of the underlying physical model. It leads to a reconstruction of the spatial structure of the gel (or paste) flow only through an analysis of the output of the simulation which is a set of unorganized points moving in a 3D space. Finally, the paper presents realistic renderings obtained by using implicit surfaces and ray-tracing techniques on the Structured Flow previously obtained.

*Keywords: Computer animation, physically-based particle models, natural phenomena, pastes and gels, visualization of an unorganized set of points.*

## 1. INTRODUCTION

There is a growing trend to model natural behaviors. These studies contribute to the understanding of the real phenomena and allow their computer synthesis. A relevant difficulty of this modeling is to determine the critical properties that induce the phenomena. *In silico* simulations are then helpful to experiment the model and to test it against reality. However, in the lack of a formal measurement of what makes two shapes or dynamics look alike, one must rely on visual inspection to validate the model [15]. The three crucial components of the modeling process are then (1) the phenomenological analysis, (2) the modeling and the simulation and (3) the visual rendering.

Multiplication of models representing a specific phenomenon raises the problem of generic representations versus "one-shot model", according the classification proposed by Fleisher [22]. This question is one of the main challenges today in computer graphics.

We believe that the physically-based particle (or mass-interaction) paradigm responds to this expectation. Simulation and visualization of models relying on this paradigm have produced animations that were sufficiently accurate for human visual recognition. It allows the reproduction of a wide range of natural phenomena or the creation of some that do not have there like in the real world. Thus, it may be seen as a general discrete formalism to express complex dynamic systems.

Conversely to this power, the simulation of such models produces only points in permanent evolution, in which the morphological information is partially lost. This lost is more substantial when the matter rearrangements are significant as in fluids, gels or pastes.

It is known that human visual perception has the capability to reconstruct this information. However, when using a mass-interaction model for computer graphics applications, we have to reconstitute the real volumetric appearance of the simulated objects. Thus, we must design methods to reconstruct volumes from such unorganized set of moving points, that conserves - rather emphasizes - the quality and the truthfulness of the dynamics.

In this paper, we first introduce a 3D model that allows the simulation of paste and gel dynamic phenomena in a realistic manner. In a second part, we describe an original visualization process, the *Flow Structuring Method*, that allows (1) the reconstruction of a morphological information from a set of unorganized points moving in a 3D space, and (2) the final rendering of this morphological information by using implicit surface and ray-tracing techniques. We then present the visual outputs we finally obtained by applying this process to the simulation data of the paste and gel model.

## 2. 3D PASTE AND GEL MODEL

### 2.1 Related Works with the particle method

In 1973, Greenspan [8] designed a single modeler to simulate all the changes of the matter states. In his formalism, the interaction law from which a pair of punctual masses interact, is composed only of a potential non-linear and non-dissipative interaction, like the well-known experimental Lenard-Jones interaction.

Numerous works have used approaching formalisms since then. The main differences between Greenspan's approach and other works are the expression of the interaction law, and the explicit introduction of a dissipative term. As examples, Terzopoulos *et al.* [17] and Tonnesen [18] used similar interaction expression, for simulation of the transition from solids to liquids; Miller and Pearce [13] explicitly added a non-linear dissipative term for animating viscous fluids.

A comparative analysis of these works was given in Luciani [12]. This article also proposed an optimal and generic model able to render the macroscopic dynamics of matter states, pursuing the model proposed by Greenspan and explicitly completed by Miller.

This model is based on the CORDIS-ANIMA general formalism, introduced by Cadoz and al. in 1990, for simulating the dynamic behaviors of physical objects by means of interacting particle networks [2, 3]. This formalism allows the expression of a wide category of non-linear interactions. Within this formalism, a non-linear interaction is represented in the more general case by a finite states automata (Figure 1), which dynamically computes at each sample of time, the values [$K_i$, $Z_i$] of the stiffness and of the viscosity terms that define a physical state [i]. Each state change is carried out by a set of conditions on the physical variables forces, displacements, velocities, etc.. The discrete expression of the interaction is then given by $F_i = K_i*D + Z_i*V$, where D is the distance between masses and V the relative velocity.

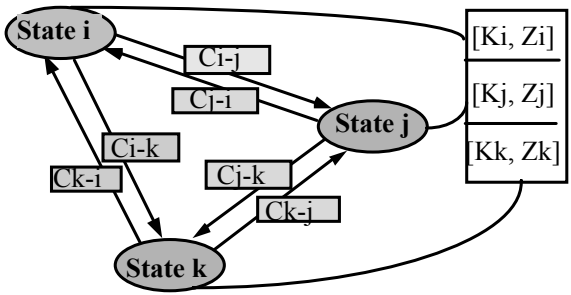

*Figure 1: Interaction defined as finite state automaton.*

As proved by numerous works in Computer Graphics as well as in computer sound generation, it is possible with this formalism to simulate a large number of complex behaviors such as dry friction, plasticity or any kind of hysteretic behaviors. As examples, Luciani *et al*. [10, 11] simulated granular materials and turbulent fluid effects with a discrete expression of a potential and dissipative non-linear interaction.

Through these works and others, physically-based particle systems have proved that they are relevant to model various objects and natural phenomena. This is a major advantage since a lot of dynamic phenomena can thus be represented within a single formalism, which is among the simplest and often does not need a precise understanding of the underlying physics of the phenomena to be modeled.

This article contributes to the demonstration of the generality of the particle approach by introducing a 3D model of pastes and gels.

## 2.2 Pastes and Gels

### 2.2.1 Pastes and gels dynamic properties

In human perception and recognition, dynamic phenomenon can be well characterized by a finite set of relevant patterns. From this point of view, paste and gel phenomena can be considered as dynamically similar.

Before packing when they strike the ground, they flow with successive chains of *curls* and *whirls*. The evolution of these curls sketches clearly the material: at the first time the flow hits the ground, a first extremum gathers. Then, it slips on soil and a new extremum gathers. Between these extrema, *points of inflection* appear, which regress along of their motion towards a new single extremum. Afterwards the two parts between these regressive extrema draw closer to create packing (fig. 2 & 3).

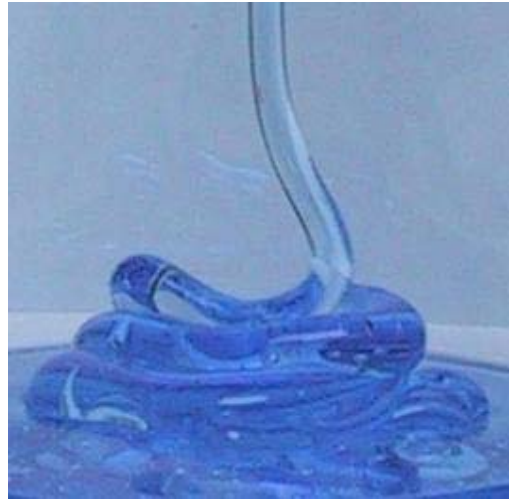

*Figure 2: Photograph of real gel*

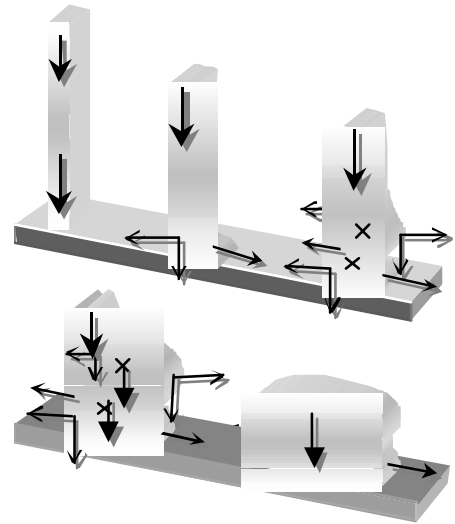

*Figure 3: Gel and Paste relevant dynamic patterns.*

### 2.2.2 The 3D model and the non-linear interaction

As said before, Luciani [12] previously proposed a generic physical particle model, composed of around 300 masses, which led to the main relevant dynamic patterns of different states of matter. Amongst other experiments, this model has been evaluated with pastes and gels simulation [12]. However, it was designed only in two dimensions.

In this paragraph, we extend this result by introducing a 3D relevant model, with the same approach, keeping in mind that, for these highly chaotic phenomena, the behavior could be deeply different in 2D and in 3D and the validation in 3D of a model designed in 2D is not *a priori* trivial.

The paste or gel model we propose is composed of a network of less than 300 masses linked with each others by n*(n-1)/2 non-linear interactions, expressed with the formalism of Cadoz & al discussed above. This non-linear interaction is both the main origin of the dynamic patterns to be generated and a relevant control for the model.

The non-linear interaction has two states, controlled by two distance thresholds: a threshold De for the potential term and a threshold Dv for the dissipative term. Behaviors are thus regulated by only four parameters : (Ke, De, Zv, Dv). The stiffness used to compute the forces is K=Ke when D<De (0, if not); and the viscosity coefficient is Z=Zv when D<Dv (0, if not).

The values of these parameters are the same for all the interactions. The model can adopt the behavior of both gels and pastes, depending on the balance between the elastic effect (Ke, De) and the damping effect (Zv, Dv). A paste effect is obtained when elasticity is preeminent, and a gel effect otherwise.

Figure 4 & 5 show some results at different times, with a very basic visualization (each mass is represented with a simple sphere).

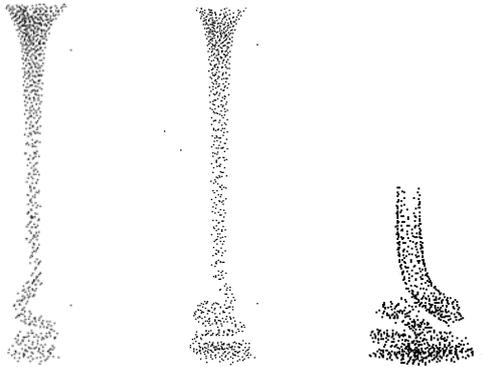

*Figure 4*: *Results of 3D paste model.*

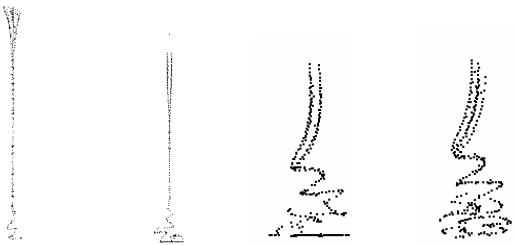

*Figure 5*: *Results of 3D gel model.*

Through this simple visualization, we can notice that the dynamic effects obtained are valid. The evolution of points corresponding to the punctual masses of the physical model is sufficient for the human recognition of the real object (gel or paste). We observe that the main relevant dynamic features of gels and pastes described before - flowing, packing, whirling - have been restituted. We consider as an important result the fact that these relevant properties were obtained in 3D with a number of masses smaller than currently thought with particle models. Only around 200 or 300 masses are used, which is equivalent to the number involved in the 2D model proposed by Luciani in 2000 [12].

## 3. SHAPE MODELING AND VISUALIZATION

In this second section, we will expose the rendering process we used in order to obtain a adequate 3D spatial extension and visualization of the 3D paste and gel model. "Adequate" means which conserves (and enhance) the main pertinent dynamic features of pastes and gels shapes evolutions as described in section 2.2.1.

### 3.1 Related works

The problem of rendering a set of moving points addresses various fields in computer graphics. As an example, some physical observations of natural phenomena (markers in fluids, scanners, etc.) provide similar punctual data. This kind of data, particle simulations results, and other fields, call for the design of rendering processes that could reconstruct the observed real objects. Several works are related to this question, both in computational geometry and computer graphics.

In computational geometry, the proposed methods usually apply to points that respect a known spatial constraint [7, 9, 14, 19]. On the contrary, in the case of the gel and paste model, spatial relations between points can change unpredictably because of the constant re-arrangement of masses in the flow.

In computer graphics, several methods were proposed in the recent years in order to render such unorganized set of point by using skeleton-based implicit surfaces. Most used techniques are *blobs* [1], *metaballs* [14] and *soft-objects* [21]. Tonnesen [18] used them to render fluids. Chiba *et al.* [4] rendered water with implicit surfaces. Subsequently, Stora *et al.* [16] combined implicit surfaces and textures to render lava, Triquet *et al.* [19] to render blood, Foster *et al.* [6] modeled and visualized water, and finally Dobashi *et al.* [5] rendered clouds. All these studies show that it is possible to visualize natural phenomena with soft and complex shapes using skeleton-based implicit surfaces.

There are two main differences between the previous approaches and the work described in this paper.

First, basically, all these works use data extracted from the underlying model that generated the set of points, or a complementary knowledge concerning the spatial organization of the set of points to be rendered. As a consequence, they may not be applied to the rendering of sets measured on real flows (through markers, for example). They are also not sufficient if the aim is to render simulated sets of points when the underlying model is not available. In the case of the present work, we suppose on the contrary that no spatial or physical information is given concerning the flow.

The second difference is that in the case of pastes and gels, if we assume that there is an implicit underlying structure that correlates the evolution of the points, we cannot use directly the points as skeletons. We need to detect and extract an organization and an evolution during flowing. The problem is not only the detection of the points that would be on a "contour", but more to specify what could be the definition of the contour. As an example, sometimes, perception will detect a fold even though points are spatially stuck.

In our case, we will extract a structural information for the flow only by using the phenomenological data, such as positions of the points, speeds, etc. The aims are then:

- To reveal, through the visualization process, some dynamic properties of the model that are present in the set of points but were not visible in the previous basic visualization;
- To allow specific rendering that emphasizes in turn various dynamic properties of the flow.
- To obtain a final realistic rendering adequate for the dynamic shapes.

## 3.2 The Flow Structuring Method

Before the rendering step in itself (which will be exposed in section 3.3), we must re-build a volume from the distributed moving points.

The *Flow Structuring Method* we propose is divided into three steps. First, it computes a *Quasi-Neighborhood Graph* with a measure of the overall distribution of the points in space throughout the simulation. Then a set of *Average Close Vectors* is build. By the end, we decide on each step and for each particle whether it is on the border of the flow or not.

### 3.2.1 Quasi-Neighborhood Graph

For each sample of a simulation, the computation of *Quasi-Neighborhood Graph* is based on a spatial relation between points: two particles are considered to be "spatially connected" and are then connected through the graph when their distance is less than a scalar $L_\alpha$.

We found that the threshold $L_\alpha$ can be chosen from an overall analysis of the data, through the relation:

$$L_\alpha = L_m / \alpha$$

where:

- $L_m$ is the average distance between all particles of the flow throughout the simulation;
- And $\alpha$ is another scalar.

The coefficient $\alpha$ represents the average capability for the matter to be stuck. It proved to be a satisfying control of the final quality of the whole rendering process, by regulating the level of details of the building volume.

We consider that a *Quasi-Neighborhood Graph* is "well-formed" when the number of the connections is necessary and sufficient to display the pertinent features of shape and dynamics and to finally to allow a correct reconstruction of the volume of the flow. As an important result, we found that, to obtain a "well-formed" graph, $\alpha$ has always to be chosen very close to 1/10. In other words, the value of $L_m$ (corrected through a coefficient $\alpha$ close to 1/10) appeared to be a relevant threshold for the construction of the *Quasi-Neighborhood Graph, whatever the simulation to be rendered can be* (paste or gel model, number of points involved, size, etc.).

Figures 6 & 7 show examples of *Quasi-Neighborhood Graphs* thus constructed.

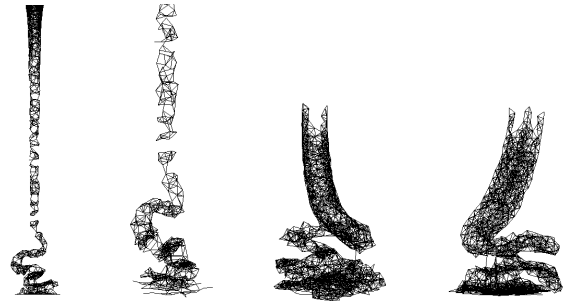

*Figure 6: Quasi-Neighborhood Graphs of paste simulation.*

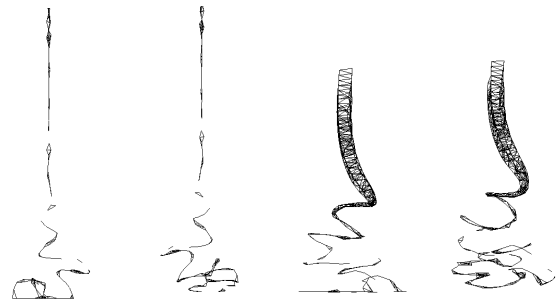

*Figure 7: Quasi-Neighborhood Graphs of gel simulation.*

The evolution of the shape is restituted in each case, with an appreciated fidelity. Relevant patterns such as extrema and packing are clearly displayed. Gel and paste shapes are differentiable. The flow of gel is thinner than that of paste. Thickening of the gel as flow goes along is rendered. The shape of the gel's flow remains helical as in a real gel. Flow's collapses in the paste are visible.

### 3.2.2 Average Close Vector

The *Quasi-Neighborhood Graph*, however does not directly lead to a surface or a volume reconstruction. As a second step, we characterize orientation neighborhood for each particle by computing an *Average Close Vector*. For a given point, we compute the average of all its connections to other points through the graph, considered as vectors.

Figures 8 & 9 display some *Average Close Vectors* we obtained. As an evidence, vectors generally point inside the flow, so that we can infer that outside lies in the opposite direction from the average close vector.

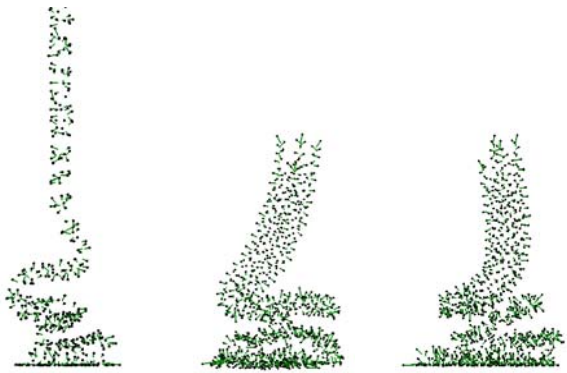

*Figure 8: Average close vectors of paste model.*

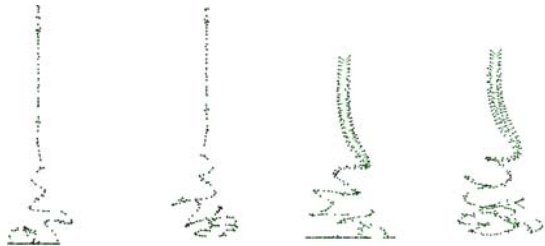

*Figure 9: Average close vectors of gel model.*

### 3.2.3 Detection of the border of the flow

Particles inside the flow have more neighbors than those outside, and a smaller average close vector. As a third step, we decide whether a point is inside the flow or on its border by analyzing the average close vector data. This decision is based on the following rule: a particle is considered on the border of the flow if the length of the neighbor vector is greater than a percentage (nearly 80%) of the length of the maximal neighbor vector.

In the figures 10 & 11, the points in black are on the detected border of the flow, whereas gray points are the inner ones. As we can see, the particles are correctly determined according to their positions in the set of points. The modeled phenomenon is still recognizable with this type of visualization.

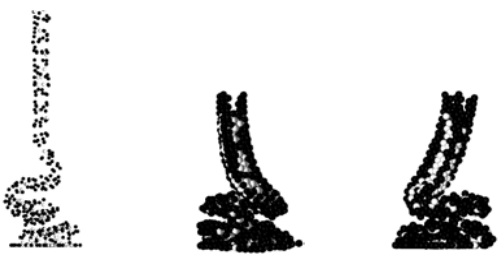

*Figure 10: Particle type determination for paste simulation.*

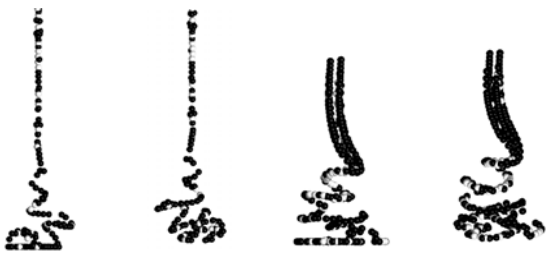

*Figure 11: Particle type determination for gel simulation.*

## 3.3 Ray-Tracing Final Rendering

To get a satisfying final rendering of the morphological structure obtained at that stage, we used implicit surfaces and ray-tracing techniques on the border points. We finally obtained, as examples, the pictures of Figure 12 and Figure 13 (see last page of the article). In these pictures, pertinent features of shape or dynamics are visible and correctly rendered.

## 4. CONCLUSION

In this paper, we introduced a single 3D physically-based particle model that allows the simulation of both the pastes and gels major dynamic features. This model extends previous results obtained in two dimensions. It is composed of a reasonable number of masses (around 300) linked one to others with homogeneous non-linear visco-elastic interactions, controlled by only four physical parameters. By adjusting these coefficients, it is possible to control the balance between the stiffness and viscosity in the resulting behavior and thus to reproduce a wide range of observed behaviors, from pastes to gels.

Beyond the quality of the dynamics it generates, this model, proves that the physically-based particle (or mass-interaction) approach is relevant for the simulation of past and gel dynamics. It is, thus, a contribution to the demonstration of the generality of this approach for the synthesis of natural phenomena. Despite the apparent complexity of the dynamic phenomena, we obtained complex dynamics patterns without a need for complex interaction such as plasticity or hysteretic property.

In the second part of the article, we introduced the *flow structuring method*, a process that allows the rendering of the outputs of the simulations, that is of unorganized sets of points. The *flow structuring method* does not use any information regarding the underlying physical model. It relies only on the phenomenological data. It analyses the spatial distribution of the points throughout the simulation to re-construct the physical structure of the flow and rebuild its volume. Three steps are used:

1. A *Quasi-Neighborhood Graph*, which produces relevant relationship between points at the morphological level;
2. A *Average Close Vectors*, which indicate where the outside of the flow is.
3. A particle's types detection, which differentiate inner particles and border particles.

At each stage of the rendering process, we verified the conformity of the results with the expected dynamic and morphological features.

As a result, we found that the average value of the distance between all the points among the simulation (corrected by a control coefficient close to 1/10) is a valid threshold for all the paste and gel simulations, regardless the number of masses involved, the kind of matter modeled (paste or gel), etc.

As a final step, we computed a skeleton-based implicit surface and ray-traced this surface. The results we obtain, such as those shown on figure 12 & 13, were, at least, promising. Examples of animations relative to this work will be found on http://www-acroe.imag.fr.

The work exposed in this paper could be extended in various directions. By taking into account others phenomenological data available, such as velocities or accelerations of each point, it should be possible to refine the *Quasi-Neighborhood Graph*, and by this way to restitute thinner and more precise morphological patterns. Another straightforward extension would be to apply the rendering method to other kinds of dynamics, such as those synthesized with particle models in the laboratory, and to determine whether or not the average distance is still a valid scalar to compute the *Quasi-Neighborhood Graph*. Experiments on sands pilling and air flow are, for example, currently performed.

**About the author**

Claire Guilbaud obtained recently his PhD in Computer Graphics from Institut National Polytechnique de Grenoble, France.

Annie Luciani is researcher and director of the ICA laboratory (Informatique et Création Artistique) from Institut National Polytechnique de grenoble and Université Joseph Fourrier, France.

Nicolas Castagné, PhD graduated, is a research engineer in ACROE (Association pour la Creation et la Recherche sur les Outils d'Expression) in Grenoble, France.

**Acknowledgements**

This work has been supported by the French Ministry of Culture. We would like to thank Peter Torvik and Pierre Boulenguez for their precious help.

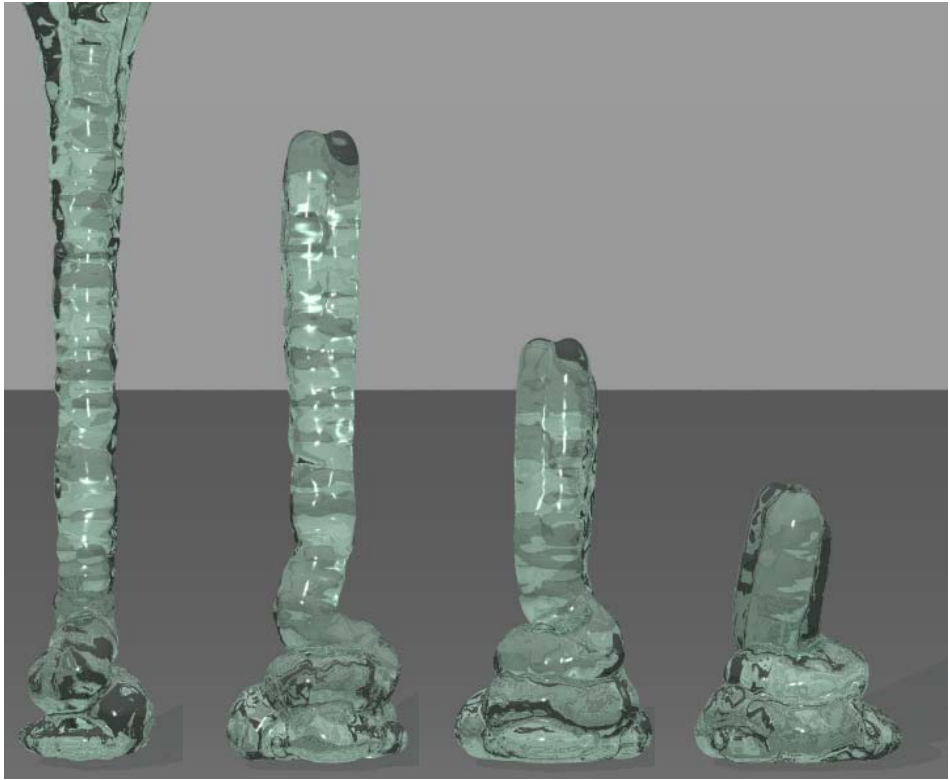

*Figure 12: Results of rendering method for paste simulation.*

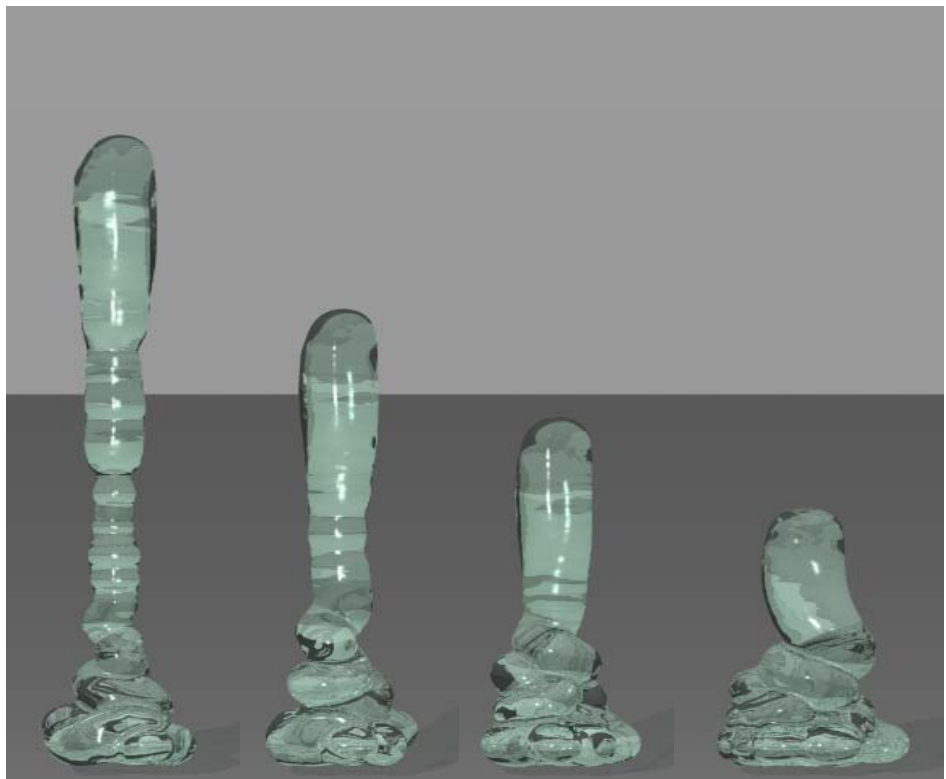

*Figure 13: Results of rendering method for gel simulation.*